\documentclass[conference, 10pt]{IEEEtran}
\IEEEoverridecommandlockouts 
\usepackage[]{graphicx}
\graphicspath{{Figures/}}
\usepackage{caption}
\usepackage{subfig} 
\usepackage{amsmath}
\usepackage{amssymb}
\usepackage{epsfig}
\usepackage{cite}
\usepackage{color}
\usepackage{balance}
\usepackage{amsmath}
\usepackage{amsfonts} % for the \checkmark command 
\usepackage{graphicx}
\usepackage{pdfpages}
\usepackage[T1]{fontenc} % For less than and greater than sign
\usepackage{array}
\usepackage{url}

\usepackage{color}
\usepackage{wrapfig}
\usepackage{lipsum}
\usepackage[hidelinks]{hyperref}
\usepackage{multirow}
\usepackage{tabularx}
\usepackage{tikz}
\usepackage{nth} % For 5th 6th
\usepackage{algpseudocode} % For algorithm 
\usepackage{algorithm,algpseudocode}
\usepackage{breqn}
\begin{document}
\title{Analyzing Solar Irradiance Variation \\From GPS and Cameras}
\author{\IEEEauthorblockN{Shilpa Manandhar$^{\dagger}$\IEEEauthorrefmark{1},
Soumyabrata Dev$^{\dagger}$\IEEEauthorrefmark{3},
Yee Hui Lee\IEEEauthorrefmark{1} and
Yu Song Meng\IEEEauthorrefmark{4}}
\IEEEauthorblockA{\IEEEauthorrefmark{1}School of Electrical and Electronic Engineering, Nanyang Technological University (NTU), Singapore}
\IEEEauthorblockA{\IEEEauthorrefmark{3}The ADAPT Centre, School of Computer Science and Statistics, Trinity College Dublin, Ireland}
\IEEEauthorblockA{\IEEEauthorrefmark{4}National Metrology Centre, Agency for Science, Technology and Research (A$^{*}$STAR), Singapore}
\thanks{$^{\dagger}$Authors contributed equally.}
\thanks{This research is funded by the Defence Science and Technology Agency (DSTA), Singapore.}
\thanks{Send correspondence to Y.\ H.\ Lee, E-mail: EYHLee@ntu.edu.sg.}
\vspace{-0.6cm}
}

% make the title area
\maketitle

\begin{abstract}
%\boldmath
The total amount of solar irradiance falling on the earth's surface is an important area of study amongst the photo-voltaic (PV) engineers and remote sensing analysts. The received solar irradiance impacts the total amount of generated solar energy. However, this generation is often hindered by the high degree of solar irradiance variability. In this paper, we study the main factors behind such variability with the assistance of Global Positioning System (GPS) and ground-based, high-resolution sky cameras. This analysis will also be helpful for understanding cloud phenomenon and other events in the earth's atmosphere.
\end{abstract}

\IEEEpeerreviewmaketitle

\section{Introduction}
The total amount of solar irradiance falling on the earth's surface is an important factor in determining the amount of generated solar energy. However, it suffers from irregular alternation of received energy because of the presence of clouds in the atmosphere. It is important to study this variability, as it directly impacts PV planning and its subsequent integration to power grid. In this context, we explain the term Cloud Radiative Effect (CRE), that is defined as the difference between the actual solar irradiance and theoretical clear-sky irradiance value. A high value of CRE means reduced solar irradiance, indicating the presence of large amount of clouds in the atmosphere. In this paper~\footnote{The source code of all simulations in this paper is available online at \url{https://github.com/Soumyabrata/irradiance-variation}}, we study this variability using GPS and sky cameras. 
 
\section{Meteorological Sensors}
In this section, we briefly describe the various meteorological sensors that are used in analyzing the cloud radiative effect on earth's surface. Section~\ref{sec:gps} discusses our methodology in computing the precipitable water vapor (PWV) from GPS. The amount of water vapor in the atmosphere impacts the received solar irradiance. We record the received solar radiance using solar sensors, which is described in Section~\ref{sec:ws}. Finally, we also compute the cloud coverage ratio using sky cameras in Section~\ref{sec:cam}.
  
\begin{figure*}[htb]
\begin{center}
\includegraphics[height=0.2\textwidth]{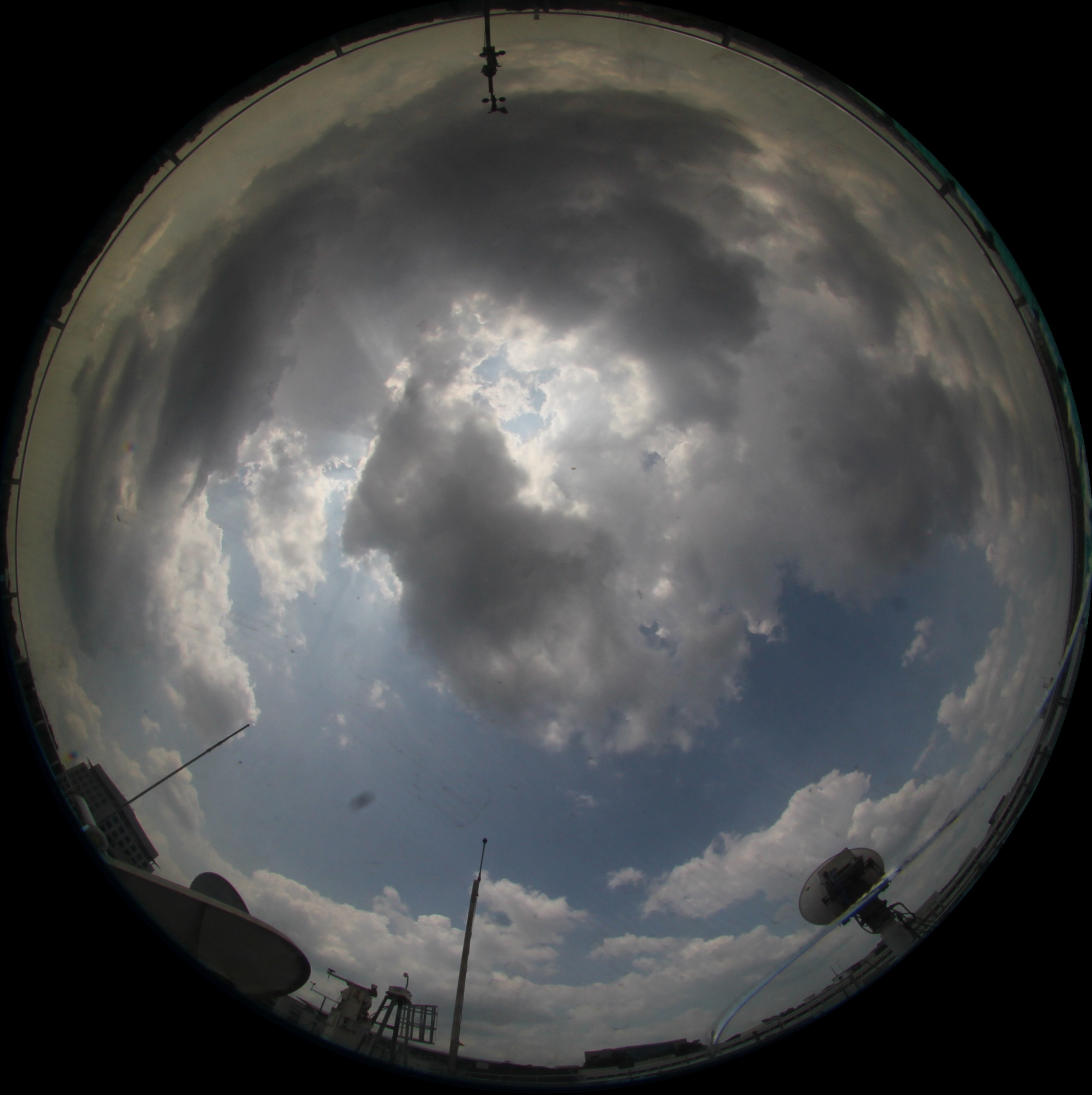}
\includegraphics[height=0.2\textwidth]{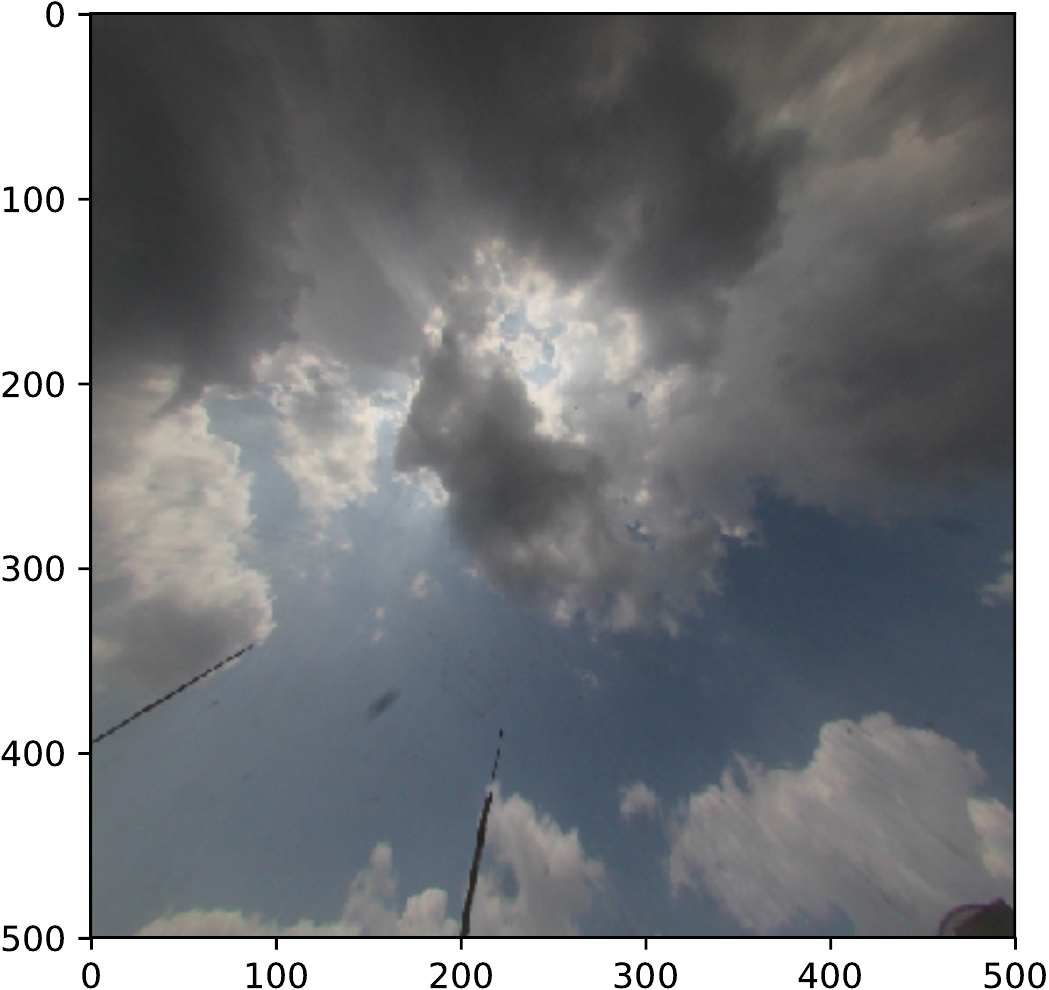}
\includegraphics[height=0.2\textwidth]{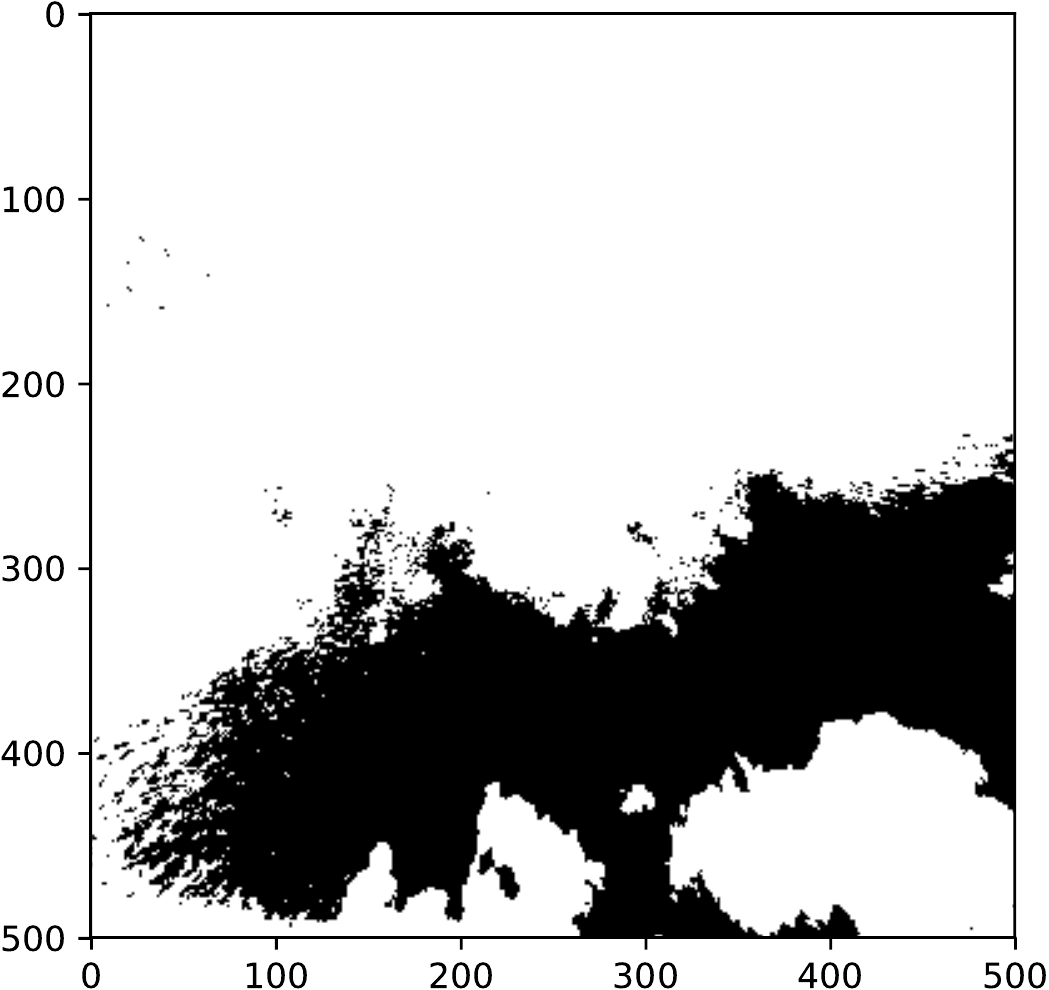}\\
\makebox[0.2\textwidth][c]{(a)}
\makebox[0.2\textwidth][c]{(b)}
\makebox[0.2\textwidth][c]{(c)}
\caption{Demonstration of a sample sky/cloud image captured using our sky camera, along with its cloud coverage computation. (a) Original image captured with our camera, (b) Undistorted image using lens calibration function~\cite{SavoyCalib}, and (c) Segmented binary image of (b) using our cloud segmentation technique~\cite{colorJSTARS}.
\label{fig:coverage-example}}
\end{center}
\end{figure*}

\subsection{Global Positioning System (GPS)}
\label{sec:gps}
The PWV values (in mm) are calculated using the zenith wet delay (ZWD), \textit{$\delta$L$_w^{o}$}, incurred by the GPS signals as shown in eq.\ref{eq1}. 

\begin{equation}
	\mbox{PWV}=\frac{PI \cdot \delta L_w^{o}}{\rho_l}
    \label{eq1}
\end{equation}     
\begin{dmath}
	PI=[-1\cdot sgn(L_{a})\cdot 1.7\cdot 10^{-5} |L_{a}|^{h_{fac}}-0.0001]\cdot cos(\frac{(DoY-28)2\pi}{365.25})+[0.165-(1.7\cdot 10^{-5})|L_{a}|^{1.65}]+f
    \label{eq2}
\end{dmath}

Here, $\rho_{l}$ is the density of liquid water (1000 kg$/m^{3}$). \textit{PI} is the dimensionless factor determined by eq. \ref{eq2}, where, \textit{L$_{a}$} is the latitude,  \textit{DoY} is day-of-year, \textit{h$_{fac}$} is 1.48 for stations from northern hemisphere and 1.25 for stations from southern hemisphere. Also, $f=-2.38\cdot 10^{-6}H$, where \textit{H} is the station height \cite{shilpaPI} and factor  \textit{f} can be ignored for stations with height less than 1000 m. In this paper, the ZWD values for a tropical IGS GPS station (IGS GPS station Id: NTUS) located at Nanyang Technological University building (1.3$^{\circ}$N, 103.68$^{\circ}$E), Singapore are processed using GIPSY OASIS software and recommended scripts \cite{GIPSY}. The PWV values are then calculated for NTUS using eq. \ref{eq1}-\ref{eq2}; with \textit{L$_{a}$} = 1.34, \textit{h$_{fac}$} = 1.48, \textit{H} = 78 m and \textit{DoY} values for all days for year 2015. The PWV values calculated has a temporal resolution of 5 minutes.

\subsection{Weather Station (WS)}
\label{sec:ws}
We perform our experiments in a rooftop of Nanyang Technological University (NTU). In addition to computing PWV values from GPS signals, we also use collocated weather station to measure the various meteorological data. We record the total rainfall, wind speed and its direction, and relative humidity. Our Davis weather station is also equipped with solar sensors that records the total incident solar irradiance on the earth's surface. All these recordings are measured at a temporal resolution of $1$ minute. 

The cloud radiative effect is calculated via the difference between total solar irradiance and the theoretical clear sky irradiance. We calculate the clear sky irradiance using Yang et al.\ clear-sky Global Horizontal
Irradiance (GHI) model~\cite{dazhi2012estimation}.

\subsection{Whole Sky Imager (WSI)}
\label{sec:cam}
Recently, with the advancement of photogrammetric techniques, remote sensing analysts have been increasingly using high-resolution, ground-based sky cameras. These cameras are popularly referred as Whole Sky Imagers (WSIs), and captures images of the sky at pre-defined interval of time. We have designed such sky cameras, and installed them collocated with the weather station. These sky cameras capture images at an interval of $2$ minutes, and provide us a localized analysis of the cloud formation. Figure~\ref{fig:coverage-example}(a) shows a sample sky/cloud image captured by our sky camera.

We compute the cloud coverage from these captured images. The image captured by our sky camera is distorted owing to the fish-eye lens of its imaging system. We undistort the fish-eye image into a $500\times500$ image, using the camera calibration function. More details on the calibration function can be found in ~\cite{SavoyCalib}. The sky/cloud pixels are estimated from the undistorted image using a ratio of \emph{red} and \emph{blue} color channels of the captured image. We employ our segmentation technique in ~\cite{colorJSTARS}, to generate the binary sky/cloud image as shown in Fig.~\ref{fig:coverage-example}. The cloud coverage ratio value is measured by computing the amount of \emph{cloud} pixels in the binary image.

\section{Relation between PWV, solar radiation and cloud coverage}
\begin{figure}[htb]
\begin{center}
\includegraphics[width=0.45\textwidth]{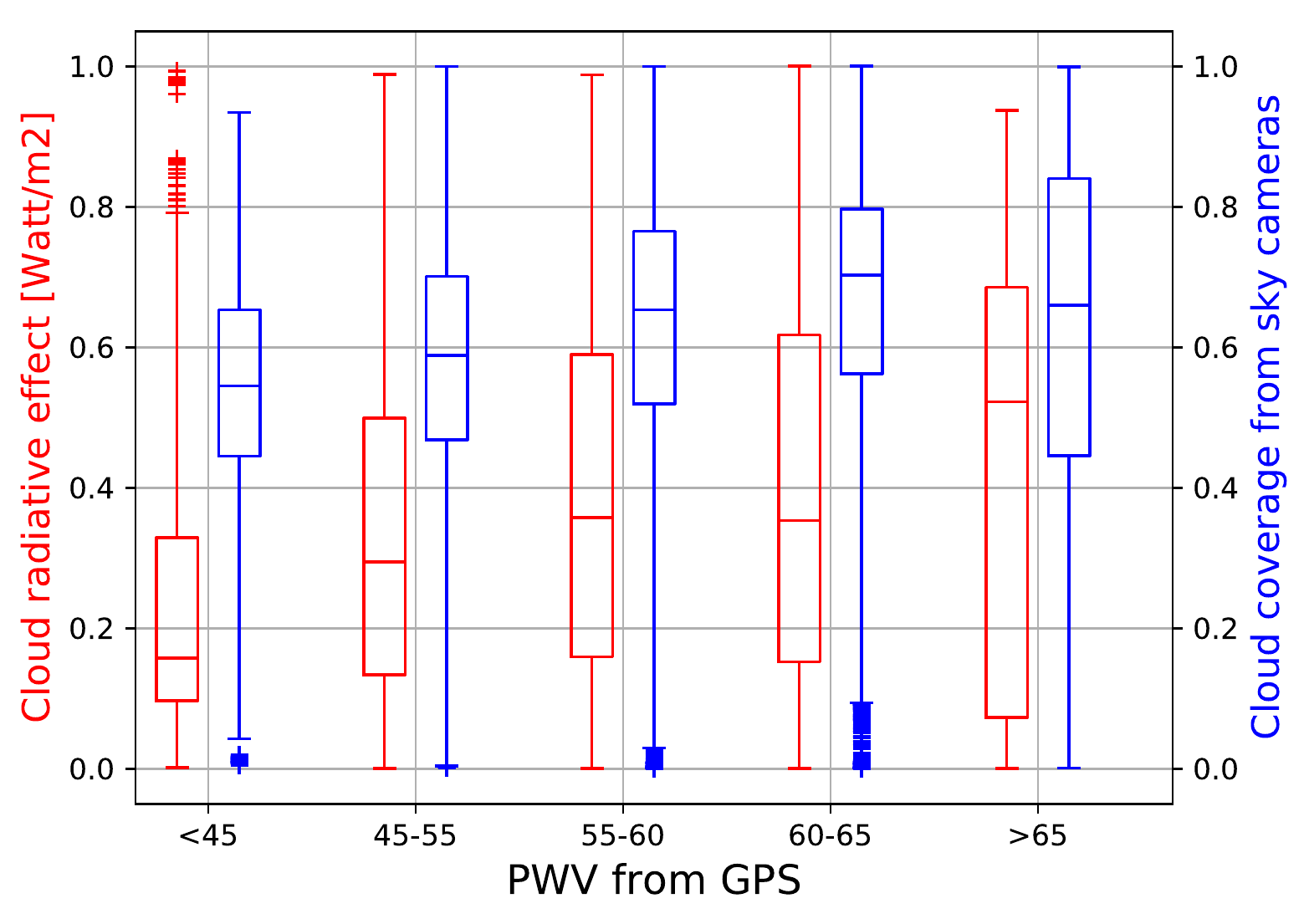}
\caption{Relationship amongst normalized solar radiation (obtained from weather station), precipitable water vapor (obtained from GPS), and normalized cloud coverage (obtained from sky cameras). It is clear that with increasing PWV values, both the net cloud radiative effect and cloud coverage increases.
\label{fig:box-plot}}
\end{center}
\vspace{-0.6cm}
\end{figure}

In our experiment, we consider all the meteorological observations for the entire year of $2015$. We consider only the day-time observations of the recordings, as solar radiation recordings are not applicable during night time. The PWV values and cloud coverage percentage are recorded for corresponding observation of cloud radiative effect. We show their relationship amongst each other via a boxplot, shown in Fig.~\ref{fig:box-plot}. We can clearly observe that the PWV increases with increasing values of CRE. This is because, an increased water content in the atmosphere indicates the presence of large amount of clouds, leading to an increased cloud radiative effect (i.e.\ large deviation of solar irradiance from clear sky radiation). Figure \ref{fig:box-plot} also indicates that the cloud coverage increases, with increasing values of CRE. This analysis provides us a framework to understand such solar irradiance variability using myriad of meteorological sensors -- GPS, sky camera and weather station.

\section{Conclusion \& Future Work}
In this paper, we studied the variability of solar irradiance using various ground-based sensors. These sensors detect the presence of clouds and water vapor in the atmosphere, which in turn, assist us in understanding the variability of solar irradiance. In our future work, we will also consider the seasonal- and diurnal- variations in our analysis.

\end{document}